\documentclass[prl,twocolumn,floatfix,superscriptaddress]{revtex4-1}
\usepackage{dcolumn,amsmath}
\usepackage{graphicx}
\usepackage{bm}
\usepackage{hyperref}

\newcommand{\Eeff}{\ensuremath{E_{\rm eff}}}
\newcommand{\eEDM}{{\em e}EDM}
\newcommand{\ecm}{\ensuremath{e {\cdotp} {\rm cm}}}

\newcommand{\de}{d_\mathrm{e}}

\begin{document}
\title{Theoretical study of HfF$^+$ cation to search for the T,P-odd interactions}

\author{L.V.\ Skripnikov}\email{leonidos239@gmail.com}
\homepage{http://www.qchem.pnpi.spb.ru}

\affiliation{National Research Centre ``Kurchatov Institute'' B.P. Konstantinov Petersburg Nuclear Physics Institute, Gatchina, Leningrad District 188300, Russia}
\affiliation{Saint Petersburg State University, 7/9 Universitetskaya nab., St. Petersburg, 199034 Russia}

\date{23.04.2017}

\begin{abstract}
The combined all-electron and two-step approach is applied to calculate the molecular parameters which are required to interpret the ongoing experiment to search for the effects of manifestation of the T,P-odd fundamental interactions in the HfF$^+$ cation by Cornell/Ye group [Science 342, 1220 (2013); J. Mol. Spectrosc. 300, 12 (2014)]. The effective electric field that is required to interpret the experiment in terms of the electron electric dipole moment is found to be 22.5 GV/cm. In Ref. [Phys. Rev. D 89, 056006 (2014)] it was shown that another source of T,P-odd interaction, the scalar-pseudoscalar nucleus-electron interaction with the dimensionless strength constant $k_{T,P}$ can dominate over the direct contribution from the electron EDM within the standard model and some of its extensions. Therefore, for the comprehensive and correct interpretation of the HfF$^+$ experiment one should also know the molecular parameter $W_{T,P}$ the value of which is reported here to be 20.1 kHz.

\end{abstract}

\maketitle

\section{Introduction}

The experiments towards the search for the effects of violation of the time-reversal and spatial parity symmetries of fundamental interactions (T,P-odd effects) such as the electron electric dipole moment (\eEDM) are prepared on several atoms and molecules containing heavy atoms. 
The experiments are of key importance to test the standard model (SM) and its extensions \cite{Commins:98,Chupp:15,Khriplovich:11}. The present best limitation on the \eEDM\ is set using the beam of neutral ThO molecules by ACME collaboration \cite{ACME:14a} and the limit is already at the level of prediction of several SM extensions~\cite{Commins:98,ACME:14a}. Cornell/Ye group has proposed to use the trapped molecular ions for such searches \cite{Meyer:06a,Meyer:08} and the corresponding experiment on the HfF$^+$ cation is in progress now \cite{Cornell:13,Cornell:14}. At present there is a number of experimental and theoretical studies of the HfF$^+$ cation \cite{Cossel:12, Cornell:13, Petrov:07a, Petrov:09b, Fleig:13, Meyer:06a, Skripnikov:08a, Le:13, Skripnikov:17b, Petrov:17a}.

Besides technical difficulties of the experiment which is conducted on the excited $^3\Delta_1$ electronic state of the cation \cite{Cossel:12} there is a common problem of all molecular experiments concerned with their interpretation in terms of fundamental parameters.
 For example, the interpretation in terms of the electron electric dipole moment requires knowledge of the parameter called the effective electric field, \Eeff\ acting on the electron EDM in a considered molecule.
This field is of the order of tens GV/cm in polar heavy atom molecules and is the most important feature of the experiments as it can lead to measurable energy shift caused by the interaction of \Eeff\ with the \eEDM. On the other hand the value of \Eeff\ cannot be measured and accurate theoretical calculations are required.
\Eeff\ for HfF$^+$ has been calculated earlier: in Ref. ~\cite{Petrov:07a} the first ab-initio calculation of  \Eeff\ was performed using the two-step method while in Ref.~\cite{Fleig:13} the 4-component approach was used. However, the experiment on HfF$^+$ under consideration can also be interpreted in terms of another source of the T,P-odd interaction  -- the scalar-pseudoscalar nucleus-electron interaction with the strength constant $k_{T,P}$.
Moreover, it is not possible to distinguish effects from the \eEDM\ and this interaction in the experiment.
As was shown in Ref.~\cite{Pospelov:14} the contribution from the scalar-pseudoscalar interaction can notably exceed that from the electron EDM within the standard model as well as within some of the extensions of the SM. For the interpretation in terms of the interaction one should use the $W_{T,P}$ molecular parameter (see below). However, according to our knowledge there are no calculations of the parameter in the literature to-date.

In the ThO experiment \cite{ACME:14a} the following limitations on the \eEDM\ ($d_e$) and $k_{T,P}$ were obtained using the experimental limitation on the energy shift (precession frequency) and calculated \Eeff\ and $W_{T,P}$ \cite{Skripnikov:13c}: $|\de|<9\times 10^{-29}$ \ecm\ and $k_{T,P} < 1.5 \cdot 10^{-8}$)~\cite{ACME:14a} (90\% confidence) (corresponding limitation on the constant $C_S$ in~\cite{ACME:14a} is recalculated here to the limitation on $k_{T,P}$ defined by Eq.(\ref{Htp}) below). Due to another charge of Hf nucleus one can expect that the relation between \Eeff\ and $W_{T,P}$ will be different in HfF$^+$ which is important for the global analysis of both experiments \cite{Chupp:15,Jung:13,Jung:14}.

The combined relativistic correlation two-step and 4-component method is applied here to compute $W_{T,P}$ as well as to recalculate \Eeff\ obtained earlier~\cite{Petrov:07a, Fleig:13} on a new level of accuracy. For the latter some important effects are considered that were missed in the previous studies.

\section{Theory}

In the Dirac-Coulomb approach the \eEDM\ effect in molecules is described by the following Hamiltonian \cite{Lindroth:89} (which can be obtained by adding a specific commutator~\cite{Lindroth:89} to the full \eEDM\ Hamiltonian~\cite{Salpeter:58}):
\begin{eqnarray}
  H_d^{{\rm eff}}= d_e\sum_j\frac{2i}{e\hbar}c\gamma^0_j\gamma_j^5\bm{p}_j^2,
 \label{Wd2}
\end{eqnarray}
where $j$ is an index over electrons, $d_e$ is the value of the \eEDM, $\gamma^0$ and $\gamma^5$ are the Dirac matrices,  and $\bm{p}$ is the momentum operator for electron.
In the case of the Dirac-Coulomb approach one can also use the \eEDM\ Hamiltonian in the following form~\cite{Lindroth:89}:
\begin{eqnarray}
  H_d^{',{\rm eff}}=2d_e\sum_j
  \left(\begin{array}{cc}
  0 & 0 \\
  0 & \bm{\sigma}_j \bm{E}_j \\
  \end{array}\right)\ ,
 \label{Wd1}
\end{eqnarray}
where $\bm{\sigma}$ are the Pauli matrices and $\bm{E}$ is the total electric field (due to the nucleus and electrons) acting on an electron.
The advantage of using Eq.~(\ref{Wd2}) is that the Hamiltonian is one-electron.
Note, however, that if the Breit interaction is considered both \eEDM\ Hamiltonians should include additional terms (see Refs.~\cite{Lindroth:89} for details and Ref.~\cite{Skripnikov:16b} for discussion).

In the molecular \eEDM\ search one always uses parameter $W_d$ or the effective electric field on electron, \Eeff. 
For this one can evaluate the expectation value of the T,P-odd operator $H_d$ (Eq.~(\ref{Wd2})):
\begin{equation}
\label{matrelem}
W_d = \frac{1}{\Omega}
\langle \Psi|\frac{H_d}{d_e}|\Psi
\rangle,
\end{equation}
where $\Psi$ is the wave function of the considered state of the molecule under consideration and
$\Omega= \langle\Psi|\bm{J}\cdot\bm{n}|\Psi\rangle$,
 $\bm{J}$ is the total electronic momentum, $\bm{n}$ is the unit vector along the molecular axis directed from Hf to F in the present case ($\Omega=1$ for the considered $^3\Delta_1$ state of HfF$^+$).
Then $E_{\rm eff}=W_d|\Omega|$.

Another source of T,P-odd interaction in HfF$^+$ is the scalar-pseudoscalar nucleus-electron interaction which does not depend on the nucleus spin. The interaction is given by the following Hamiltonian (see \cite{Ginges:04}, Eq.~(130)):
\begin{eqnarray}
  H_{T,P}=i\frac{G_F}{\sqrt{2}}Zk_{T,P}\sum_j \gamma^0_{j}\gamma^5_{j}\rho_N(\textbf{r}_j),
 \label{Htp}
\end{eqnarray}
where $k_{T,P}$ is the dimensionless constant of the interaction, $G_F$ is the Fermi-coupling constant and $\rho_N(\textbf{r})$ is the nuclear density normalized to unity ($k_{T,P}$ and $\rho_N(\textbf{r}_j)$ in Eq.~(\ref{Htp}) correspond to $C_p^{SP}$ and $\rho(\textbf{r})$ in Eq.~(130) of Ref.~\cite{Ginges:04}, respectively).
The interaction is characterized by the molecular parameter $W_{T,P}$ which is required to extract the strength constant $k_{T,P}$ from the experimental data:
\begin{equation}
\label{WTP}
W_{T,P} = \frac{1}{\Omega}
\langle \Psi|\frac{H_{T,P}}{k_{T,P}}|\Psi
\rangle.
\end{equation}

Note that \Eeff\ and $W_{T,P}$ parameters cannot be measured (since the \eEDM\ and $k_{T,P}$ are unknown) and have to be obtained from an accurate molecular electronic structure calculation.

According to the ``Atoms-In-Compounds'' concept ~\cite{Skripnikov:15b,Titov:14a,Zaitsevskii:16a} there is an indirect possibility to estimate the accuracy of calculated \Eeff\ and $W_{T,P}$ parameters as they are mainly determined by the behavior of a valence wave function in the region close to the heavy atom nucleus. 
Such parameters are called  Atoms-In-Compounds (AIC) properties or characteristics.
With a good accuracy the AIC characteristics are localized on a heavy atom and do not depend on the bonding electronic density in contrast to some other types of properties \cite{Mayer:07,Sizova:08,Sizova:08b,Sizova:09}.
As demonstrated in Refs.~\cite{Skripnikov:15b,Titov:14a,Zaitsevskii:16a} they are determined by so-called W-reduced density matrix with a very small dimension and the corresponding matrix elements. Thus, if one of the AIC property is measured one can estimate the uncertainty of other calculated properties.
Therefore, one often calculates the magnetic dipole hyperfine structure (HFS) constant of heavy atom which is the AIC property which can be measured. Note, however, that HFS constant is determined by the diagonal matrix elements of W-reduced density matrix while \Eeff\ is determined by the off-diagonal ones~\cite{Skripnikov:15b}.

The HFS constant $A_{||}$ is defined by the following matrix element:
\begin{equation}
 \label{Apar}
A_{||}=\frac{\mu_{\rm Hf}}{I\Omega}
   \langle
   \Psi|\sum_i\left(\frac{\bm{\alpha}_i\times
\bm{r}_i}{r_i^3}\right)
_z|\Psi
   \rangle, \\
\end{equation}
where $\mu_{\rm Hf}$ is the magnetic moment of an isotope of the Hf nucleus having the spin $I$.
Here we use  $\mu_{\rm Hf}$=0.7936 $\mu_N$ and I=$3.5$ for  $^{177}$Hf.

\section{Electronic structure calculation details}

The HfF$^+$ cation contains 80 electrons and has the following electron configuration of the considered first excited electronic state $^3\Delta_1$ (in the naive ionic model): $[\dots] 6s^1 5d^1$ for Hf and $1s^2 2s^2 2p^6$ for F. 

In Refs.\cite{Skripnikov:16b,Skripnikov:17a} a very accurate computation scheme of calculation of the AIC properties in heavy atoms and molecules containing heavy atoms has been developed.
The scheme combines the direct 4-component Dirac-Coulomb(-Breit) approach and the two-step approach \cite{Titov:06amin,Skripnikov:15b,Skripnikov:16a} developed earlier by our group.
The latter approach divides the whole calculation into two stages. (I) calculation of the valence and outer-core part of the molecular wave function within the generalized relativistic effective core potential (GRECP) method \cite{Titov:99,Mosyagin:10a,Mosyagin:16}.
(II) Restoration of the correct 4-component behavior of the valence and outer core wave function in the spatial core region of a heavy atom using the nonvariational procedure developed in Refs.~\cite{Titov:06amin,Skripnikov:15b,Skripnikov:16a,Skripnikov:11a}. Note, that the procedure is rather universal and has been recently extended to the case of three-dimensional periodic structures (crystals) in Ref.~\cite{Skripnikov:16a}. GRECP and the restoration procedure were also successfully used for precise investigation of different diatomics~\cite{Lee:13a,Skripnikov:15b,Skripnikov:14c,Petrov:13,Kudashov:13,Kudashov:14,Skripnikov:09,
Skripnikov:15d,Petrov:14,Skripnikov:13c,Skripnikov:09a,Skripnikov:08a,
Skripnikov:15c,Skripnikov:14a}.

Such division in the two-step approach leads to a number of computational savings with respect to the basis set used (see e.g. \cite{Titov:06amin,Skripnikov:15b} for details). As a result it is possible to treat high order correlation effects at lower cost using compact basis sets \cite{Skripnikov:13a} and, on the other hand, consider additional corrections on very large basis sets using the scalar-relativistic (1-component) variants of the GRECP operator \cite{Skripnikov:16b}.
Combination of the two-step method with 4-component relativistic all-electron approaches allows one to treat the contribution to the considered parameters from the correlation of the inner-core electrons (which are excluded from the correlation treatment in the current formulation of the GRECP method) and exclude approximations introduced in the nonvariational procedure due to  treatment of the valence approximation to the full GRECP operator.

Here the combined method similar to that employed in Refs.~\cite{Skripnikov:16b,Skripnikov:17b} is applied and includes the following steps.
(I) Calculation of the the main correlation contributions within the 52-electron coupled cluster with single, double and noniterative triple cluster amplitudes, CCSD(T), theory using the 4-component Dirac-Coulomb Hamiltonian.
Here the uncontracted CVQZ basis set for Hf \cite{Dyall:07,Dyall:12}, [34,30,19,13,4,2], and the aug-ccpVQZ basis set \cite{Dunning:89,Kendall:92} with two removed g-type basis functions for F, [13,7,4,3]/(6,5,4,3), were used. 
For the correlation calculation virtual spinors were truncated at energy of 50 Hartree.
(II) Calculation of the correlation contribution from the $1s..3d$ inner-core electrons of Hf excluded in step (I). The contribution was evaluated at the CCSD level as the difference between the values of the calculated parameters with correlation of all 80 electrons of HfF$^+$ and with 52 electrons as in stage (I).
The CVDZ \cite{Dyall:07,Dyall:12} basis set on Hf and the cc-pVDZ~\cite{Dunning:89,Kendall:92} 
 basis set on F were used. Virtual spinors were truncated at energy of 7000 Hartree (see Ref.\cite{Skripnikov:17a} where the required cutoff was investigated).
(III) Calculation of the correction on the Gaunt interaction.
Note that this correction is only an order of magnitude estimate in the case of \Eeff\ (see Ref.\cite{Skripnikov:16b} for detailed discussion). In this treatment the version of the Hartree-Fock method which is optimal only for the $^3\Delta_1$ state was used in contrast to a treatment within the average-of-configuration Hartree-Fock method used in all other calculations.
(IV) Calculation of high-order correlation effects correction as the difference of the calculated values obtained at the level of the coupled cluster with single, double, triple and noniterative quadruple amplitudes method, CCSDT(Q), and the CCSD(T) method.
20 valence electrons were correlated using the 2-component (with spin-orbit effects included) two-step approach within the 12-electron semilocal version of the GRECP operator for Hf used earlier in Refs.\cite{Petrov:07a,Petrov:09b,Skripnikov:08a}.
The reduced version of the basis set which was used in Refs.~\cite{Petrov:07a,Petrov:09b,Skripnikov:08a} has been applied for Hf, [12,16,16,10,8]/(6,5,5,1,1) while the ANO-I basis set \cite{Roos:05} reduced to [14,9,4,3]/(4,3,1) was used for fluorine. The basis will be denoted as CBas.
(V) Calculation of the basis set correction for 52 valence and outer electrons of HfF$^+$ within the scalar-relativistic two-step approach within the semilocal version of the 44-electron GRECP operator~\cite{Mosyagin:10a,Mosyagin:16}. Here the influence of the additional 7 $g-$, 6 $h-$ and 5 $i-$ basis functions on Hf 
was considered. For this two basis sets were used for Hf. Lbas: [20,20,20,15,4,2] and LbasExt: [20,20,20,15,11,8,5].

Finally, the total of parameter $X$ ($X=$\Eeff, $W_{T,P}$ and $A_{||}$) of the excited $^3\Delta_1$ state of  HfF$^+$ were calculated using the following expression:
\begin{equation}
\begin{array}{l}
X({\rm TOTAL}) = X(\mbox{52e-4c-CCSD(T), QZ})  \\
                  \\   
                 + X(\mbox{80e-4c-CCSD(T), DZ}) -\\
                  X(\mbox{52e-4c-CCSD(T), DZ}) \\
                  \\
                 + X(\mbox{4c-Dirac-Fock-Gaunt, QZ}) - \\
                  X(\mbox{4c-Dirac-Fock, QZ}) \\
                  \\
                 + X(\mbox{two-step-2c-20e-CCSDT(Q),CBas}) - \\ 
                 X(\mbox{two-step-2c-20e-CCSD(T),CBas}) \\
\\
                 + X(\mbox{two-step-1c-52e-CCSD(T),LBasExt}) - \\
                  X(\mbox{two-step-1c-52e-CCSD(T),LBas}) \\
\end{array}
 \label{Calc}
\end{equation}

The Hf$-$F potential curve and internuclear distance in the $^3\Delta_1$ state have been calculated by us earlier and found to be in a very  good agreement with the  experiment \cite{Cossel:12}. It was set here to be  3.41 Bohr.
Hartree-Fock and integral transformations calculations were done using the {\sc dirac12} code \cite{DIRAC12}. Relativistic coupled cluster calculations were performed within the {\sc mrcc} code  \cite{MRCC2013}. Scalar-relativistic calculations were performed using the {\sc cfour} code \cite{CFOUR,Gauss:91,Gauss:93,Stanton:97}. 
Matrix elements of operators (\ref{Wd2}), (\ref{Htp}) and (\ref{Apar}) over molecular bispinors were calculated using the code developed by us in Ref.~\cite{Skripnikov:16b}.
The code to perform restoration of the correct 4-component structure of valence and outer core wavefunction in the core region of heavy atom which was used here was developed by us in Ref.~\cite{Skripnikov:15b}.

\section{Results and discussions}

Results of calculations are given in Table \ref{TResults}.

\begin{table}[!h]
\centering
\caption{
Calculated values of the effective electric field (\Eeff), parameter of the scalar-pseudoscalar nucleus-electron interaction ($W_{T,P}$) and hyperfine structure constant (A$_{||}$) of the $^3\Delta_1$ state of HfF$^+$ using the coupled cluster method compared to previous calculations.
}
\label{TResults}
\begin{tabular}{llrrr}
\hline\hline
Ref. & Method                  & \Eeff & $W_{T,P}$ & A$_{||}$                                  \\
                      &              & (GV/cm)  & (kHz)        &(MHz) \\
                    
\hline    

\cite{Petrov:07a} &  20e-SODCI$^a$    & 24.2  & ---    &  -1239 \\
\cite{Fleig:13}       &  VTZ/34e-MR-CISD+T$^b$    & 23.3  & ---    &  --- \\

\hline
                   
\hline
This            &CVQZ/52e-4c-CCSD(T)$^c$   & 22.5   & 19.8      & -1375   \\
work     &                                 &        &           &       \\   
            &Inner-core contribution$^c$       & 0.7   & 0.6       & -42       \\
       &Gaunt correction$^c$                   & -0.7   & -0.3      & +2       \\                   
     &High-order correlation effects,          & 0.00   & 0.0       & 0      \\                    
     &~~~CCSDT(Q)-CCSD(T)$^d$                     &        &           &       \\        
       &Basis set correction$^d$               & 0.0    & 0.0       & -13     \\

\\
                    &Total         & 22.5   & 20.1     & -1429 \\
    \hline\hline

\end{tabular}
\\
\flushleft
$^a$ SODCI - Spin-orbit direct configuration interaction method, see Ref.\cite{Petrov:07a} for details. \\
$^b$ See Ref. \cite{Fleig:13} for details and designations.\\
$^c$ Calculated within the 4-component approach.\\
$^d$ Calculated within the two-step GRECP/Restoration approach.
\end{table} 

In Refs. \cite{Skripnikov:15a,Skripnikov:16b} a very detailed comparison between different methods to treat electron correlation effects including multireference configuration interaction methods used in previous studies of HfF$^+$ \cite{Petrov:07a,Fleig:13} has been performed. It was found that the combined procedure that uses coupled cluster theory (see Eq.~\ref{Calc}) and applied here gives the most stable and reliable results.

The contribution to all the considered AIC characteristics from the $1s..3d$ inner-core electrons of Hf (3\%, see Table~\ref{TResults}) is not negligible though it has not been considered earlier in Refs.~\cite{Petrov:07a, Fleig:13}. Note, that it is almost equal for all of the considered properties (see Table~\ref{TResults}). Interestingly, this contribution is close also to that found in Ref.~\cite{Skripnikov:17a} for the T,P-odd constant $R_s$ for the francium atom (2\%).
In Ref.~\cite{Skripnikov:17a} is was shown that the mechanism of this contribution is mainly due to spin polarization of the core electrons, therefore for parameters such as \Eeff, $W_{T,P}$, A$_{||}$ which are mainly determined by $s-$ and $p-$ electrons of heavy atom one can use rather small basis sets if sufficient number of $s-$ and $p-$ basis functions are included whereas the number of basis functions for higher-order harmonics can be safely reduced. This is the case for the used CVDZ basis set of Hf in calculation of the contribution.

It was shown in Ref.\cite{Skripnikov:16b} that the contribution of the high-order correlation effects to the value of \Eeff\ can be calculated within the two-step procedure with high accuracy. It is confirmed in the present calculations: contribution of noniterative triple cluster amplitudes is very close within the 4-component and two-step approaches (about -0.4 GV/cm).

The result for $W_{T,P}$ and \Eeff\ for HfF$^+$ can be compared with those of ThO molecule \cite{Skripnikov:16b} ($W_{T,P}$(ThO)$=113.1$ kHz, \Eeff(ThO)$=79.9$ GV/cm) calculated within the same combined method used here. The relation of \Eeff\ and $W_{T,P}$ for the two molecules is not the same as expected due to considerably different charges of Hf and Th nuclei.

As was noted in Ref.~\cite{Skripnikov:16b} the most important part of the uncertainty of \Eeff\ (not of $W_{T,P}$) is due to the approximate treatment of the Breit interaction. If this interaction is considered the \eEDM\ Hamiltonian (\ref{Wd2}) should be modified to include some two-electron operator \cite{Lindroth:89} which is not possible in the current consideration \cite{Skripnikov:16b}.
As an additional test the Gaunt contribution to \Eeff\ was also estimated within the EDM Hamiltonian (\ref{Wd1}) in approximation when one neglects electron contribution to the electric field $\bm{E}$. The resulted contribution to \Eeff\ is -0.3 GV/cm which is smaller than that calculated within the Hamiltonian (\ref{Wd2}) similar to the ThO case~\cite{Skripnikov:16b}.
We include the whole estimated contribution to \Eeff\ from the Breit interaction to the final uncertainty of \Eeff. As can be seen from Table \ref{TResults} the other uncertainties are smaller.
Therefore, the final uncertainty of the presented T,P-odd constants is estimated to be less than 4\%.
Up to now there is no experimental value of A$_{||}$ constant for the $^{177}$HfF$^+$ cation.

Note that in the literature there are different definitions of the $k_{T,P}$ constant and $W_{T,P}$ parameter.
For example, in Ref.~\cite{ACME:14a} one used the $W_S$ constant that is connected with $W_{T,P}$ by the following relation:
\begin{equation}
 \label{WsWtp}
W_S = W_{T,P} \cdot (Z+N) / Z,
\end{equation}
where $N$ is the number of neutrons in the considered nucleus. Thus, for the case of $^{180}$HfF$^+$: $W_S$=50.3 kHz.

The final values of \Eeff\ and $W_{T,P}$ given in Table~\ref{TResults} are recommended for interpretation on the ongoing experiment on HfF$^+$ \cite{Cornell:17} in terms of fundamental quantities of the \eEDM\ and $k_{T,P}$.

\section*{Acknowledgement}
This work is supported by the RFBR, according to the research project No.~16-32-60013 mol\_a\_dk.


\end{document}